# Thermal rectification and negative differential thermal conductivity based on a parallel-coupled double quantum-dot

Yanchao Zhang[1,*], Shanhe Su[2,†]

[1] *School of Science, Guangxi University of Science and Technology, Liuzhou 545006, People's Republic of China*

[2] *Department of Physics, Xiamen University, Xiamen 361005, People's Republic of China*

We investigate the heat flow transport properties of a parallel-coupled double quantum-dot system connected to two reservoirs with a temperature bias in the Coulomb blockade regime. We demonstrate that the effects of thermal rectification and negative differential thermal conductance (NDTC) exist in this system and analyze the influences of energy level difference and Coulomb interaction on the thermal rectification and NDTC. We find that this system can achieve a high thermal rectification ratio and NDTC when the asymmetry factor of the system is enhanced.

[*] Email: zhangyanchao@gxust.edu.cn
[†] Email: shanhesu@xmu.edu.cn

## I. INTRODUCTION

Controlling heat flow at the nanoscale have attracted significant attention in the fields of modern science and technology because of fundamental and potential applications for the development of nanoscale thermal devices [1-5]. Recent studies have demonstrated that the effects of thermal rectification and negative differential thermal conductance (NDTC) are the two most important features for building the basic components of functional thermal devices, which are the key tools for the implementation of solid-state thermal circuits [6, 7]. Thermal rectification is a phenomenon in which heat transport depends on the direction of the flow, and is first discovered in one-dimensional nonlinear lattice [8]. Based on different microscopic mechanisms, a very significant rectifying effect was exhibited [6] and subsequently demonstrated experimentally by Chang *et al.* using nanotubes [9]. Li and coworkers put forward the first theoretical proposal for a thermal transistor and clarified that NDTC is the crucial element for operation of the thermal transistor [7]. In the following years, thermal rectification and NDTC have been extensively studied in different systems both theoretically [10-22] and experimentally [23-25]. At the same time, various thermal transistor models have also been proposed, such as near-field thermal transistors [26], far-field thermal transistors [27-29], electrochemical thermal transistor [30], and quantum thermal transistors [31, 32]. Moreover, new concepts for thermal devices, such as thermal logical gates [33-35], thermal memory [36-39], thermal memristor [40], and heat circulator [41], have also been proposed and demonstrated.

In recent years, double quantum-dot systems with two-terminal or multi-terminal structure have been regarded as new candidates in the designs of thermoelectricity [42-50], thermal rectification [51-55], and logical stochastic resonance [56]. In addition, related applications have been extended to the field of thermometry [57, 58] and quantum information [59-61]. These advances suggest that in the near future thermal currents could be manipulated flexibly as electron circuits today [51]. For a double quantum-dot system with two-terminal, when the quantum dots are capacitively coupled in series, there is only exchange heat but no electron transport through the system [42, 51]. Therefore, Ruokola *et al.* first applied this feature to introduce a single-electron thermal diode consisting of two quantum dots or metallic islands coupled to two reservoirs and explored the rectification performance of thermal diode [51]. And then, Aligia *et al.* systematically studied the heat flow transport properties through two capacitively coupled quantum dots in the spinless case [62]. On the other hand, two quantum dots can connected in parallel to two reservoirs, which provides an excellent platform for studying the interaction, interference effect [63-65], and thermoelectricity (see e.g., Refs. [66-68] and references therein). Recently, Sierra *et al.* investigated the nonequilibrium transport properties of a double quantum-dot system connected in parallel to two reservoirs, and found that negative differential conductance can take place by changing the thermal gradient [67].

In this paper, we further investigate the nonequilibrium transport properties of a parallel-coupled double quantum-dot system connected to two reservoirs via asymmetric energy-depending tunnel barriers. We will focus on the heat flow transport properties of this system. We demonstrate that the phenomenon of thermal rectification effect and NDTC can occur in this system and analyze the influences of energy level difference and Coulomb interaction on the thermal rectification effect and NDTC. This paper is organized as follows. In Sec. II, the model and basic physical theory of a parallel-coupled double quantum-dot system are briefly described. In Sec. III, the thermal rectification effect is investigated. In Sec. IV, the influences of the energy level difference and Coulomb interaction on the NDTC are discussed. Finally, in Sec. V, the main results are summarized.

## II. MODEL AND THEORY

We consider a double quantum-dot system connected to left ($L$) and right ($R$) reservoirs with two energy-dependent tunneling channels in parallel structure, as shown schematically in Fig. 1. The left ($L$) and right ($R$) reservoirs are characterized by different temperatures $T_L$ and $T_R$. The quantum dots $QD_u$ and $QD_d$ with single energy $\varepsilon_u$ and $\varepsilon_d$ are Coulombic coupled to each other and interact only through the long-range Coulomb force such that they can only exchange energy but no particles.

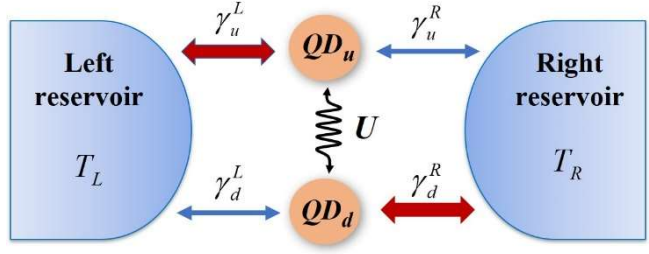

FIG. 1. Model of a parallel-coupled double quantum-dot system connected to two reservoirs in the Coulomb blockade regime. $U$ is the long-range Coulomb interaction and describes the exchanged energy which can be transferred from one dot to the other dot. The energy-dependent tunneling rates are indicated as $\gamma_u^L$, $\gamma_u^R$, $\gamma_d^L$, and $\gamma_d^R$.

The quantum states of double quantum-dot system is denoted by the charge configuration $|n_u, n_d\rangle$, where $n_\alpha$ ($\alpha = u, d$) is the occupation number of quantum dot $\alpha$. In the Coulomb blockade regime, the double quantum-dot system is well described by four quantum states labeled as $|0,0\rangle$, $|1,0\rangle$, $|0,1\rangle$, and $|1,1\rangle$, where 0 or 1 represents that the sites of $QD_u$ and $QD_d$ is empty or filled, respectively. The occupation probabilities for four quantum states are given by the diagonal elements of the density matrix, $\boldsymbol{p} = (p_{00}, p_{10}, p_{01}, p_{11})^T$. In the limit of weak tunneling coupling ($\hbar\gamma \ll k_B T$), the broadening of energy levels can be neglected and the transmission through a tunnel barrier is well described by sequential tunneling of a single electron [42]. The off-diagonal density matrix elements do not contribute to steady state transport and can be neglected. Thus, the time evolution of occupation probabilities is given by a master equation. The matrix form can be written as $d\boldsymbol{p}/dt = \boldsymbol{Mp}$, where $\boldsymbol{M}$ denotes the matrix containing the transition rates and is given by Fermi's golden rule, i.e.,

$$M = \begin{pmatrix} -\sum_{\alpha,v}\Gamma_{\alpha 0}^{v+} & \sum_{v}\Gamma_{u0}^{v-} & \sum_{v}\Gamma_{d0}^{v-} & 0 \\ \sum_{v}\Gamma_{u0}^{v+} & -\sum_{v}\Gamma_{u0}^{v-}-\sum_{v}\Gamma_{d1}^{v+} & 0 & \sum_{v}\Gamma_{d1}^{v-} \\ \sum_{v}\Gamma_{d0}^{v+} & 0 & -\sum_{v}\Gamma_{d0}^{v-}-\sum_{v}\Gamma_{u1}^{v+} & \sum_{v}\Gamma_{u1}^{v-} \\ 0 & \sum_{v}\Gamma_{d1}^{v+} & \sum_{v}\Gamma_{u1}^{v+} & -\sum_{\alpha,v}\Gamma_{\alpha 1}^{v-} \end{pmatrix}. \quad (1)$$

where $\Gamma_{\alpha n}^{v\pm} = \gamma_{\alpha n}^{v} f_{v}^{\pm}(\varepsilon_{\alpha}+U_{\alpha n})$ describe tunneling rates that take an electron into (+) or out (-) of the quantum dot $\alpha$ through reservoir $v$ ($v = L, R$) when the occupation number of the other quantum dot is $n$. $\gamma_{\alpha n}^{v}$ is the bare tunneling rate, $f_{v}^{+}(x) = [1+e^{(x-\mu_v)/k_B T_v}]^{-1}$ is the Fermi function and $f_{v}^{-}(x) = 1 - f_{v}^{+}(x)$. $U_{\alpha n}$ is the charging energy of quantum dot $\alpha$, which depending on the occupation number $n$ of the other quantum dot (see Appendix). $\varepsilon_{\alpha}$ is the bare energy of the single level in the quantum dot $\alpha$.

The steady state probabilities are solved by the stationary solution of the master equation, i.e., $Mp = 0$, together with the normalization condition $\sum p = 1$. Thus, the total steady-state charge current from the left reservoir to the right reservoir is given by

$$I_L = I_d^L + I_u^L, \quad (2)$$

where $I_d^L$ and $I_u^L$ are the charge currents through the $QD_d$ and $QD_u$, respectively, and are given as

$$I_d^L = q\sum_n \left(\Gamma_{dn}^{L+} p_{n0} - \Gamma_{dn}^{L-} p_{n1}\right), \quad (3)$$

and

$$I_u^L = q\sum_n \left(\Gamma_{un}^{L+} p_{0n} - \Gamma_{un}^{L-} p_{1n}\right), \quad (4)$$

where $q$ is elementary positive charge. Total steady-state charge current from the right reservoir to the left reservoir is $I_R = I_d^R + I_u^R$. At steady-state, $I_L + I_R = 0$.

Then the steady-state heat flow from the left reservoir to the double quantum-dot system is given by

$$J_L = J_d^L + J_u^L. \tag{5}$$

with

$$J_d^L = \sum_n \left( \varepsilon_d + E_d^L + \delta_{1n} U \right) \left( \Gamma_{dn}^{L+} p_{n0} - \Gamma_{dn}^{L-} p_{n1} \right) \tag{6}$$

and

$$J_u^L = \sum_n \left( \varepsilon_u + E_u^L + \delta_{1n} U \right) \left( \Gamma_{un}^{L+} p_{0n} - \Gamma_{un}^{L-} p_{1n} \right), \tag{7}$$

which give the heat flow through $QD_d$ and $QD_u$, respectively. In Eqs. (6) and (7), $E_d^L = U_{d0} - \mu_L$ and $E_u^L = U_{u0} - \mu_L$ (see Appendix). At steady-state, $J \equiv J_L = J_R$, where $J_R = J_d^R + J_u^R$ is the heat flow from the double quantum-dot system to the right reservoir.

In the following discussion, we set $T_L = T + \Delta T/2$ and $T_R = T - \Delta T/2$, where $T = (T_L + T_R)/2$ is the average temperature of the two reservoirs and $\Delta T = T_L - T_R$ is the temperature bias. The single energy $\varepsilon_u = \varepsilon + \Delta\varepsilon/2$ and $\varepsilon_d = \varepsilon - \Delta\varepsilon/2$ with $\Delta\varepsilon = \varepsilon_u - \varepsilon_d$ being the energy level difference. Without loss of generality, we set the $\mu_\nu = 0$ and $\varepsilon = 0$ as the energy reference point. We define $\Delta T > 0$, that is, the left reservoir is at higher temperature, which is the forward configuration with the heat spontaneous flowing from the left reservoir to the right reservoir. In the reverse configuration when $\Delta T < 0$, the right reservoir is at higher temperature and the heat will flow from the right reservoir to the left reservoir. The asymmetric energy-dependent transport barriers are defined by the bare tunneling rate $\gamma_{\beta n}^\nu = \gamma$, except $\gamma_{dn}^L = \gamma_{un}^R = \lambda\gamma$, where $0 \leq \lambda \leq 1$ describes the symmetry factor for the coupling of the quantum dots with the two heat reservoirs. When $\lambda = 0$ means that $QD_u$ is only coupled to the left reservoir and $QD_d$ is only coupled to the right reservoir. In this case our model reverts to two interacting quantum dots coupled in series with two reservoirs at different temperatures [51, 62]. Since electrons cannot be transferred between $QD_u$ and $QD_d$, it is clear that the particle current is zero in this case. When $\lambda = 1$, each quantum dot has exactly the same left and right tunneling rate, corresponding to the case of perfect symmetry.

**III. THERMAL RECTIFICATION EFFECT**

We consider the parallel-coupled double quantum-dot system that is in contact with two heat reservoirs maintained at a temperature bias $\Delta T$, and eventually reaches a nonequilibrium steady state characterized by a steady-state heat flow $J$. When the system is symmetric, one can be predicted that $J(-\Delta T) = -J(\Delta T)$. In this case, reversing the temperature bias only changes the direction of the heat flow. Conversely, if the system is asymmetric, we expect that $J(-\Delta T) \neq -J(\Delta T)$, that is, thermal rectification effect exists. This asymmetric configuration is the origin of the thermal rectification effect.

The contour plot of the heat flow as functions of the temperature bias $\Delta T$ and energy level difference $\Delta \varepsilon$ is shown in Fig. 2 (a) for $U/\hbar\gamma = 10$. It is found that the heat flow is asymmetric with respect to the separatrix $\Delta \varepsilon = 0$, which clearly demonstrates the thermal rectification effect. In the regimes of $\Delta \varepsilon > 0$ and $\Delta \varepsilon < 0$, the rectification effect is completely anti-symmetric. This means that different rectifying directions can be obtained by changing the positive and negative values of $\Delta \varepsilon$. Moreover, within a certain range, the rectification effect is obviously enhanced with the increase of $|\Delta \varepsilon|$. In Fig. 2 (b), we plot the heat flow as functions of the temperature bias $\Delta T$ and Coulomb interaction $U$ for $\Delta \varepsilon/\hbar\gamma = 20$. As can be seen, with the increase of Coulomb interaction $U$, both forward and reverse heat flows show a trend of first increasing and then decreasing. Moreover, the thermal rectification effect is obvious in the small Coulomb interaction range.

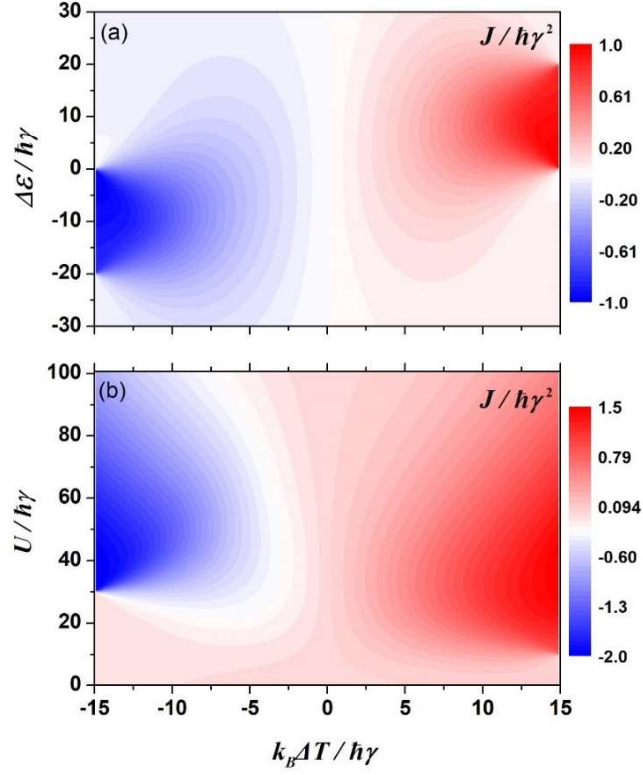

FIG. 2. Heat flow in the parallel-coupled double quantum-dot system. (a) Heat flow as functions of the temperature bias and energy level difference for $U/\hbar\gamma = 10$. (b) Heat flow as functions of the temperature bias and Coulomb interaction for $\Delta\varepsilon/\hbar\gamma = 20$. Other parameters: $\lambda = 0.01$, $q^2/C_\alpha = 20\hbar\gamma$, and $k_B T = 7.5\hbar\gamma$.

The heat flow as a function of the temperature bias $\Delta T$ for different values of the asymmetric factor $\lambda$ is shown in Fig. 3. It is clearly seen that when $\Delta T > 0$ (forward configuration) the heat flow increases with the increase of $\Delta T$, while in the region $\Delta T < 0$ (reverse configuration) the heat flow is suppressed. The heat flow in the reverse configuration is significantly lower than that in forward configuration. In this way, the model acts like a good thermal conductor in the forward configuration while in the reversed way, the model acts like a thermal insulator, which demonstrates clearly the thermal rectification effect. The thermal rectification effect becomes significant as the symmetric factor $\lambda$ decreases (asymmetry enhancement).

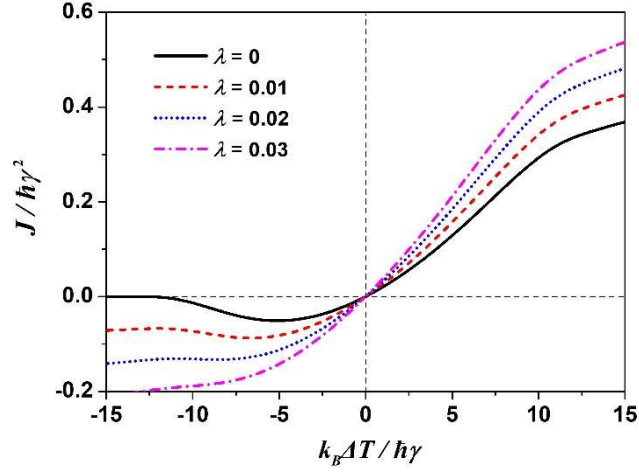

FIG. 3. Heat flow as a function of the temperature bias for different values of the asymmetric factor $\lambda$. Parameters: $\Delta\varepsilon/\hbar\gamma = 20$, $U/\hbar\gamma = 10$, $q^2/C_\alpha = 20\hbar\gamma$, and $k_B T = 7.5\hbar\gamma$.

In order to quantitatively describe the efficiency of thermal rectification, we introduce two main parameters. One is thermal rectification ratio $\mathcal{R}$ which is defined as the ratio of the heat flow $J_f$ in the forward configuration to the heat flow $J_r$ in the reverse configuration, i.e.,

$$\mathcal{R} = \left|\frac{J_f}{J_r}\right|. \tag{8}$$

This parameter indicates that $\mathcal{R} = 1$ is the case of no rectification, while $\mathcal{R} \to \infty$ means a perfect thermal diode. The other is thermal rectification coefficient expressed as the sum of $J_f$ and $J_r$ divided by the difference of these two flows [69, 70], i.e.,

$$\mathcal{C} = \left|\frac{J_f + J_r}{J_f - J_r}\right|. \tag{9}$$

It is zero, $\mathcal{C} = 0$, when there is no rectification and takes the value $\mathcal{C} = 1$ for a perfect thermal diode.

In Figs. 4(a) and 4(b), the resulting thermal rectification ratio $\mathcal{R}$ and thermal rectification coefficient $\mathcal{C}$ versus the temperature bias $\Delta T$ are depicted for different symmetric factor $\lambda$. It is found that the thermal rectification ratio is always greater than 1, i.e., $\mathcal{R} > 1$ and increases significantly with the symmetric factor decreases. A considerable rectification ratio can be obtained when the symmetric factor $\lambda \to 0$. In the case of $\lambda = 0$, our model recovers to an ideal single-electron heat diode [51]. To visualize this rectification effect more clearly, the thermal rectification coefficient $\mathcal{C}$ is shown in Fig. 4(b). As can be seen, with the decrease of $\lambda$, the thermal rectification coefficient increases gradually, which indicates that the rectification effect is enhanced gradually. In particular, the thermal rectification coefficient $\mathcal{C} = 1$ when the temperature bias $|\Delta T| > 12$ in the case of $\lambda = 0$, which means an ideal thermal diode.

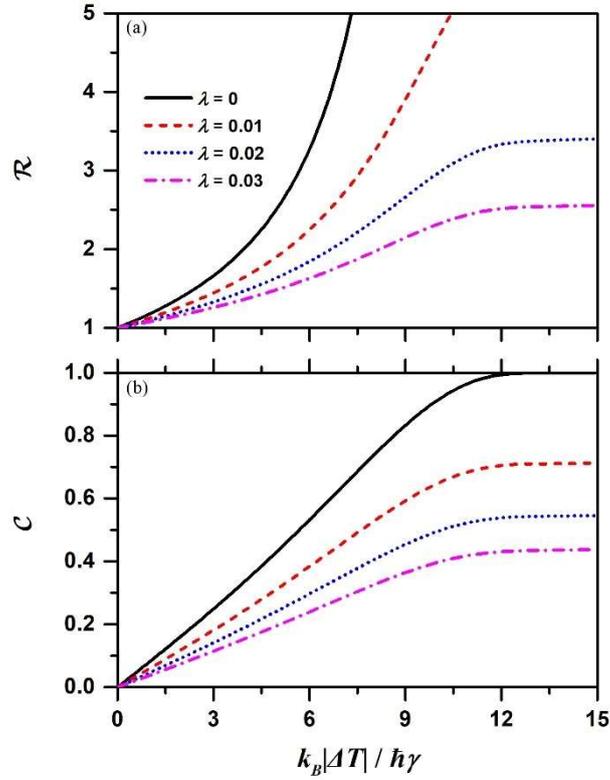

FIG. 4. Heat rectification ratio (a) and thermal rectification coefficient (b) as a function of the temperature bias for different values of the symmetric factor $\lambda$. The values of parameters are the same as those used in Fig. 3.

## IV. NEGATIVE DIFFERENTIAL THERMAL CONDUCTANCE

By examining the behavior of the heat flow curves in Fig. 3 again, we can also see that the heat flow decreases with the increase of the temperature bias in the area of reverse temperature bias. This implies the negative differential thermal conductance (NDTC) effect, i.e., the differential thermal conductance (DTC) $\partial J/\partial \Delta T < 0$. In order to observe to how the asymmetric factor $\lambda$ influences the DTC, Fig. 5 shows the variation curves of the DTC with the temperature bias under different values of the asymmetric factor $\lambda$. It can be seen that there does exhibit the NDTC in the area of reverse temperature bias, although the region of NDTC decreases and finally disappears with increasing the asymmetric factor $\lambda$.

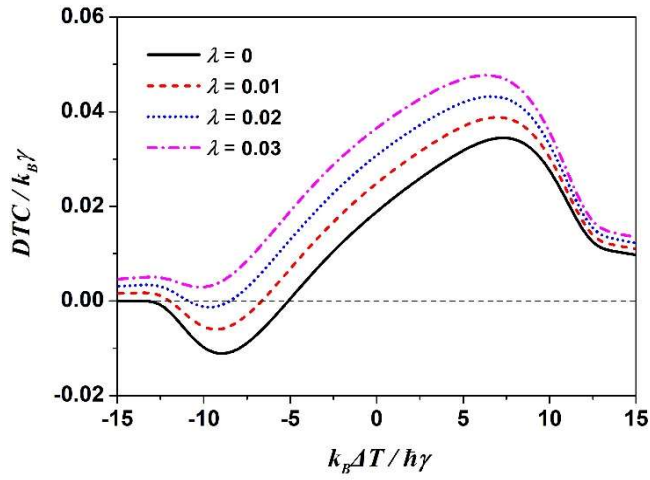

FIG. 5. Differential thermal conductivity as a function of the temperature bias for different values of the asymmetric factor $\lambda$. The values of parameters are the same as those used in Fig. 3.

Fig. 6 further reveal the influences of the energy level difference and the Coulomb interaction on the performance of the NDTC. Therefore, only the region of NDTC is shown in Fig. 6. It is found from Fig. 6(a) that when $\Delta\varepsilon > 0$, As $\Delta\varepsilon$ increases, the NDTC appears first at $\Delta T < 0$ and then at $\Delta T > 0$, however, is reversed at $\Delta\varepsilon < 0$. Fig. 6(b) shows that with the increase of Coulomb interaction $U$, the NDTC can also appear in the two regions of $\Delta T < 0$ and $\Delta T > 0$, and the region of NDTC is larger in $\Delta T < 0$. In Fig. 6, the contour lines of $\partial J/\partial \Delta T = 0$ and $\partial J/\partial \Delta T = -0.03$ are given. They respectively indicate the region of NDTC and its variation trend. Note that the NDTC can reach $-1.45$ although not shown in Fig. 6(a).

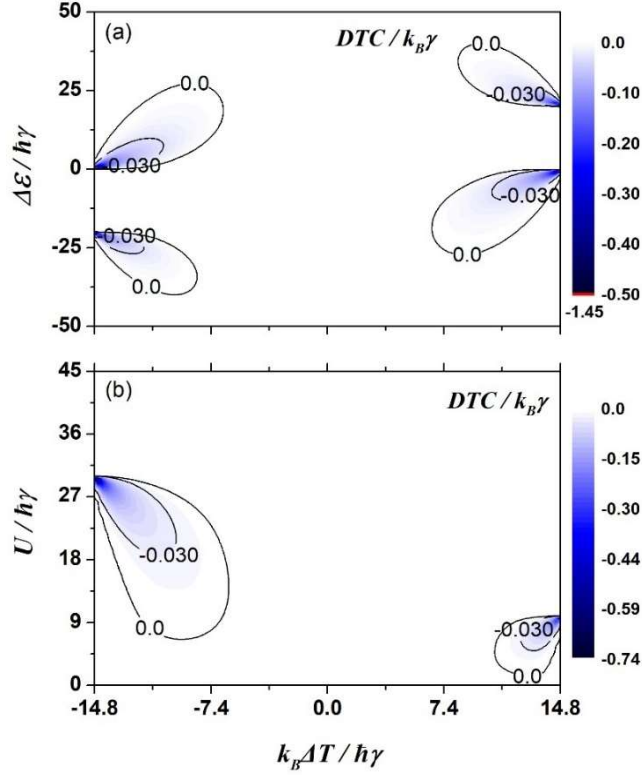

FIG. 6. Differential thermal conductivity in the parallel-coupled double quantum-dot system. (a) Differential thermal conductivity varying with the temperature bias and energy level difference for $U/\hbar\gamma = 10$. (b) Differential thermal conductivity varying with the temperature bias and Coulomb interaction for $\Delta\varepsilon/\hbar\gamma = 20$. Other parameters: $\lambda = 0.01$, $q^2/C_\alpha = 20\hbar\gamma$, and $k_B T = 7.5\hbar\gamma$.

## V. CONCLUSIONS

We have demonstrated that the negative differential thermal conductance and the thermal rectification effect can take place in a parallel-coupled double quantum-dot system. The influences of energy level difference and Coulomb interaction on the NDTC and thermal rectification effect are carefully analyzed. An ideal thermal diode and significant NDTC can be obtained by a careful fine-tuning of the system's parameters, i.e., the asymmetric factor $\lambda$. We found that our system can achieve a high thermal rectification ratio when the asymmetry of the system is enhanced (that is, $\lambda$ decreases). These results can stimulate further theoretical and experimental research in the field of nanoscale heat transport and may open up potential applications for thermal rectification at nanoscales.

**ACKNOWLEDGEMENTS**

This paper is supported by the National Natural Science Foundation of China (No. 11947010, No. 11805159), and the Science and Technology Base and Talent Project of Guangxi (No. AD19110104).

**APPENDIX: CHARGING ENERGIES OF A PARALLEL-COUPLED DOUBLE QUANTUM-DOT SYSTEM**

The equivalent circuit of a parallel-coupled double quantum-dot system connected to two reservoirs is sketched in Fig. 7, where the quantum dot $\alpha$ are coupled to their respective reservoirs $v$ via the tunneling junction with junction resistance and capacitance $C_\alpha^v$ in parallel. $C_0$ is the capacitance between the $QD_u$ and the $QD_d$. $V_\alpha^v = \mu_\alpha/q$ determines the bias voltage of the reservoirs $v$ connected to quantum dot $\alpha$, where $\mu_\alpha$ is the chemical potential and $q$ is elementary positive charge. $V_\alpha$ is the electrostatic potential of quantum dot $\alpha$.

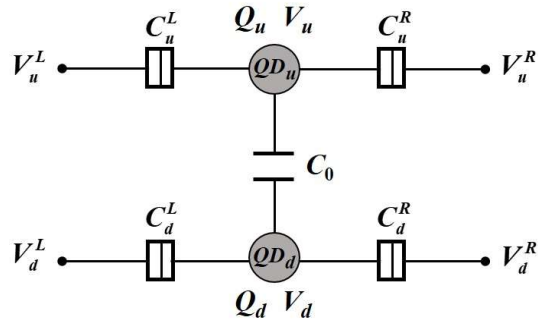

FIG. 7. The equivalent circuit of a parallel-coupled double quantum-dot system connected to two reservoirs.

The total charges of $QD_u$ and $QD_d$ are given by the sum of the charges on all of the capacitors connected to $QD_u$ and $QD_d$, i.e.,

$$Q_u = C_u^L \left( V_u - V_u^L \right) + C_u^R \left( V_u - V_u^R \right) + C_0 \left( V_u - V_d \right), \tag{A1}$$

$$Q_d = C_d^L \left( V_d - V_d^L \right) + C_d^R \left( V_d - V_d^R \right) + C_0 \left( V_d - V_u \right). \tag{A2}$$

These two equations can be expressed more compactly in a matrix form

$$\begin{pmatrix} Q_u + C_u^L V_u^L + C_u^R V_u^R \\ Q_d + C_d^L V_d^L + C_d^R V_d^R \end{pmatrix} = \begin{pmatrix} C_{\Sigma u} & -C_0 \\ -C_0 & C_{\Sigma d} \end{pmatrix} \begin{pmatrix} V_u \\ V_d \end{pmatrix}. \tag{A3}$$

where $C_{\Sigma u} = V_u^L + V_u^R + C_0$ and $C_{\Sigma d} = V_d^L + V_d^R + C_0$ define the total capacitances of $QD_u$ and $QD_d$. The quantum dot electrostatic potentials $V_u$ and $V_d$ are then expressed by using the capacitance matrix

$$\begin{pmatrix} V_u \\ V_d \end{pmatrix} = \frac{1}{C_{\Sigma u} C_{\Sigma d} - C_0^2} \begin{pmatrix} C_{\Sigma u} & C_0 \\ C_0 & C_{\Sigma d} \end{pmatrix} \begin{pmatrix} Q_u + C_u^L V_u^L + C_u^R V_u^R \\ Q_d + C_d^L V_d^L + C_d^R V_d^R \end{pmatrix}. \tag{A5}$$

Thus, the electrostatic energy for the double quantum dot system is given by

$$U(n_u, n_d) = \frac{1}{2} \begin{pmatrix} Q_u + C_u^L V_u^L + C_u^R V_u^R \\ Q_d + C_d^L V_d^L + C_d^R V_d^R \end{pmatrix}^T \begin{pmatrix} V_u \\ V_d \end{pmatrix}. \tag{A6}$$

where $Q_u = n_u q$ and $Q_d = n_d q$. $n_u$ and $n_d$ are electron numbers of $QD_u$ and $QD_d$ respectively. In the Coulomb blockade regime, each of quantum dots can be occupied only by zero or one electron ($n_{u/d} = 0,1$). Thus, the electrostatic energies for the four quantum states read $U(0,0)$, $U(1,0)$, $U(0,1)$, and $U(1,1)$. Thus the charging energy of quantum dot $\alpha$ is defined by $U_{\alpha n}$, depending on whether the other quantum dot is empty $(n=0)$ or occupied $(n=1)$. When an electron tunnels into a quantum dot while the other dot is empty, the charging energies of $QD_u$ and $QD_d$ are, respectively, given by

$$U_{u0} = U(1,0) - U(0,0), \tag{A7}$$

$$U_{d0} = U(0,1) - U(0,0). \tag{A8}$$

However, when the other quantum dot is occupied, the charging energies are, respectively, given by

$$U_{u1} = U(1,1) - U(0,1) = U_{u0} + U, \tag{A9}$$

$$U_{d1} = U(1,1) - U(1,0) = U_{d0} + U. \tag{A10}$$

where

$$U = \frac{q^2 C_0}{C_{\Sigma u} C_{\Sigma d} - C_0^2}. \tag{A11}$$

is the exchanged energy which can be transferred from one dot to the other dot. Here, we consider the case $V_u^L = V_d^L = V_L$, $V_u^R = V_d^R = V_R$, and $C_u^L = C_u^R = C_d^L = C_d^R \equiv C_\alpha$. Thus one has

$$U = \frac{q^2 C_0}{4 C_\alpha (C_\alpha + C_0)}, \tag{A12}$$

and

$$E_u^L = E_d^L = \frac{q^2 (2C_\alpha + C_0) + 4q C_\alpha (C_\alpha + C_0)(V_R - V_L)}{8 C_\alpha (C_\alpha + C_0)}, \tag{A13}$$

$$E_u^R = E_d^R = \frac{q^2 (2C_\alpha + C_0) - 4q C_\alpha (C_\alpha + C_0)(V_R - V_L)}{8 C_\alpha (C_\alpha + C_0)}. \tag{A14}$$

where $E_u^L = U_{u0} - \mu_L$, $E_u^R = U_{u0} - \mu_R$, $E_d^L = U_{d0} - \mu_L$, and $E_d^R = U_{d0} - \mu_R$.